\documentclass[showpacs,aps,prl,twocolumn,superscriptaddress,10pt]{revtex4-2}
\usepackage{graphicx} % Include figure files
\usepackage{bm}% bold math
\usepackage{color}

\usepackage{amsmath}
\usepackage{amssymb}
\usepackage{enumerate}
\usepackage{xspace}
\usepackage{mathrsfs}
\usepackage{mathptmx}
\usepackage[utf8]{inputenc}

\newcommand{\sS}{\ensuremath{\mathbf{S}}\xspace}

\newcommand{\smmu}{\ensuremath{\mu_{\mathrm{s}}}\xspace}

\newcommand{\smB}{\ensuremath{\mathbf{B}}\xspace}

\newcommand{\sms}{\ensuremath{\mathbf{S}}\xspace}
\newcommand{\Jij}{\ensuremath{J_{ij}}\xspace}
\newcommand{\Jnn}{\ensuremath{J_{ij}^{\mathrm{nn}}}\xspace}
\newcommand{\Jnnn}{\ensuremath{J_{ij}^{\mathrm{nnn}}}\xspace}

\newcommand{\IrMn}{\ensuremath{\text{IrMn}_3}\xspace}
\newcommand{\Neel}{N\'eel\xspace}
\newcommand{\Lonetwo}{L1$_2$\xspace}

\newcommand{\new}[1]{\textcolor{black}{#1}}

\begin{document}

\title{The atomistic origin of exchange anisotropy in non-collinear IrMn/CoFe}
%\title{The natural model of exchange bias}
%in IrMn/CoFe}

\author{Sarah Jenkins}
\email{sarah.jenkins@york.ac.uk}
\affiliation{Department of Physics, University of York, York, YO10 5DD, UK}
\author{Wei Jia Fan}
\affiliation{Shanghai Key Laboratory of Special Artificial Microstructure Materials and Technology and Pohl Institute of Solid State Physics and School of Physics Science and Engineering, Tongji University, Shanghai 200092, China}
\author{Roxana Gaina}
\affiliation{University of Fribourg, Department of Physics and Fribourg Center for Nanomaterials, Chemin du Musee 3, CH-1700 Fribourg, Switzerland}
\affiliation{Laboratory for Neutron Scattering and Imaging, Paul Scherrer Institut, CH-5232 Villigen PSI, Switzerland}
\author{Roy.~W.~Chantrell}
\affiliation{Department of Physics, University of York, York, YO10 5DD, UK}
\author{Timothy Klemmer}
\affiliation{Seagate Technology, Fremont, California 94538, USA}
\author{Richard~F.~L.~Evans}
\email{richard.evans@york.ac.uk}
\affiliation{Department of Physics, University of York, York, YO10 5DD, UK}

\begin{abstract}
%The exchange interaction determines the ferromagnetic (FM) or antiferromagnetic (AFM) ordering of atomic spins~\cite{neel1}. 
%When ferromagnets and antiferromagnets are coupled together they exhibit the exchange bias effect, where a unidirectional interface exchange field causes a shift of the magnetic hysteresis loop. The effective magnitude of this interface exchange field is at most a few percent of the bulk exchange arising from pinned interfacial spins in the antiferromagnet. 
Anti-ferromagnetic spintronic devices could offer ultra fast dynamics and a higher data density than conventional ferromagnetic devices. One of the challenges designing such devices is the control and detection of the magnetisation of the anti-ferromagnet due to its lack of stray fields, and this is often achieved through the exchange bias effect. In exchange biased systems the pinned spins are known to comprise a small fraction of the total number of interface spins, yet their exact nature and physical origin has so far been elusive. Here we show that in the technologically important disordered $\gamma$ \IrMn/CoFe structure the pinned spins arise from the small imbalance in the number of spins in each magnetic sublattice in the antiferromagnet due to the naturally occurring atomic disorder. These pinned spins are strongly coupled to the bulk antiferromagnet explaining their stability. Moreover, we find that the ferromagnet strongly distorts the interface spin structure of the antiferromagnet, causing a large reversible interface magnetisation that does not contribute to exchange bias but does increase the coercivity. We find that the uncompensated spins are not localised spins which occur due to point defects or domain walls but instead constitute a small motion of every AFM spin at the interface. This unexpected finding resolves one of the long standing puzzles of exchange bias and provides a route to developing optimized nanoscale antiferromagnetic spintronic devices.
\end{abstract}

%\pacs{75.10.Hk,75.20.-g,75.50.Ss,75.60.Jk,75.78.Jp}
\maketitle

\textit{Introduction} - Antiferromagnetic spintronics is a rapidly developing field using the naturally fast dynamics and zero magnetic moment to store, transmit and manipulate information~\cite{BaltzRMP2018,JungwirthNnano2016}. While pure electrical stimulation and detection of antiferromagnets is possible~\cite{GodinhoNatComm2018,ZeleznyNatPhys2018,AsaJAP2020}, nanoscale manipulation, stability and control is often best achieved through coupling to an adjoining ferromagnet via the exchange bias effect~\cite{FukamiNatMat2016,LinNatMat2019,KimAPL2019}. The exchange bias effect occurs when a ferromagnet (FM) is coupled to an antiferromagnet (AFM), causing a shift of the magnetic hysteresis loop ~\cite{MBean,Berkowitz,EBreview,HOhldag}. Despite the ubiquity of exchange biased devices and extensive measurements~\cite{EBreview,HOhldag,bulkspins} understanding the microscopic origins of the exchange bias effect is the preserve of models and has so far proved elusive for non-collinear antiferromagnets such as PtMn and IrMn \cite{SchullerJMMM2016}.

Many theoretical models have been proposed to explain the exchange bias effect~\cite{domainstate,spinflop,mechanisms,model,Mauri1987SimpleSubstrate}. The first model came from Meikeljohn and Bean, where they assumed a perfectly uncompensated spin structure at the interface, giving predicted values for the exchange bias field  orders of magnitude larger than those obtained from their experimental measurements~\cite{MBean}. To reduce the discrepancy, over the next 60 years models were proposed which assumed that the lowest energy magnetic configuration may not be a perfectly rigid AFM and a perfectly uniform FM. Most of these models were based on the idea of AFM domains. The first model to utilise AFM domains came from Mauri \textit{et al}~\cite{Mauri1987SimpleSubstrate}, assuming a perfectly flat interface and a perfectly compensated spin structure. He proposed that the formation of domain walls reduced the predicted exchange bias by an order of magnitude to match the experimental results. The model accurately calculated the exchange bias shift but fails to predict the observed increase in coercivity. The first micromagnetic model of exchange bias came from Koon \textit{et al}~\cite{Koon1997CalculationsInterfaces},  modelling a perfectly flat uncompensated interface. The uncompensated spins occur because in the interface region the spins are frustrated by the AFM and FM exchange coupling, causing canting of the AFM spins. This turns out to be possible only by forming a stable domain wall in the AFM while keeping the FM mostly aligned with the canted moment. The domain state model of Nowak \textit{et al}~\cite{Nowak2000DomainTheory} assumed that the AFM is perfectly compensated, but uncompensated spins occur due to domains in the AFM. The fundamental limitation with the domain state model is that the anisotropy constant necessary to create domains in the AFM is orders of magnitude higher than measured experimentally\cite{EBreview}. So far these theoretical models have been unable to provide a robust mechanism for explaining the exchange bias effect in non-collinear antiferromagnetic systems with realistic spin structures.

Here we present a natural atomistic model of exchange bias applied to $\gamma$-\IrMn / CoFe bilayers, including a realistic $3Q$ tetrahedral spin structure and intrinsic atomic disorder in IrMn. Our model gives accurate values for the exchange bias loop shift and the increase in coercivity without the need for AFM domains or grain boundaries. We find that in IrMn systems the exchange bias originates from the natural structural disorder, creating a small statistical imbalance in the number of interfacial spins. These spins are pinned by exchange coupling to the antiferromagnet, yet are accompanied by a reversible component, manifesting as a distortion of the interface spin structure. Our calculations unambiguously identify the physical nature of these spin components, finally resolving one of the most complex physical effects in nature.

\textit{Method} - Our simulations were performed using an atomistic spin model with the \textsc{vampire} software package~\cite{vampire}. The energetics of the system is described by the spin Hamiltonian
\begin{eqnarray}
\mathscr{H} &=& -\sum_{i<j} \Jij \sS_i \cdot \sS_j - \frac{k_N}{2} \sum_{i \neq j}^z (\mathbf{S}_i \cdot  \mathbf{e}_{ij})^2 \nonumber \\
&& - \sum_i k_{\mathrm{u}} (\sS \cdot \mathbf{e}_z)^2 - \sum_i \smmu \sms_i \cdot \smB,
\label{eq:hamiltonian}
\end{eqnarray}
with $\sS_i$ describing the spin direction on site $i$, $k_N = -4.22 \times 10^{-22}$ is the \Neel pair anisotropy constant and $\mathbf{e}_{ij}$ is a unit vector from site $i$ to site $j$, $z$ is the number of nearest neighbours and \Jij is the exchange interaction. The effective exchange interactions (\Jij) were limited to nearest ($\Jnn = -6.4 \times 10^{-21}$ J/link) and next nearest ($\Jnnn = 5.1 \times 10^{-21}$ J/link) neighbours~\cite{FS,Jenkins2019MagneticIrMn3}. The CoFe is simulated with a nearest neighbour approximation and a weak easy-plane anisotropy $k_{\mathrm{u}}$ to simulate the effects of the demagnetising field of a thin film. The exchange coupling across the FM/AFM interface is set at 1/5th of the bulk exchange values as calculated by \textit{ab-initio} methods~\cite{abinitio2}.

Spin Dynamics simulations were done solving the stochastic Landau-Lifshitz-Gilbert equation with a Heun numerical scheme~\cite{Garcia-Palacios1998Langevin-dynamicsParticles}. Our model naturally reproduces the low temperature ground state spin structures where the ordered alloy forms a triangular (T1) spin structure with an angle of 120 degrees between adjacent spins and the disordered alloy forms a tetrahedral (3Q) spin structure with 109.5 degrees between spins \cite{FS} in agreement with previous neutron scattering experiments~\cite{neutrons,kohn} and theoretical calculations~\cite{abinitio,theory,MC}. \new{These ground state spin structures are shown in Fig. 1 (c).} The simulations also reproduce different \Neel ordering temperatures for the ordered and disordered phases in close agreement with experimental values of 730 K~\cite{AFM} and 960K~\cite{Tomeno} due to different degrees of spin frustration.
\begin{figure}[!tb]
\centering
\includegraphics[width= 0.5\textwidth]{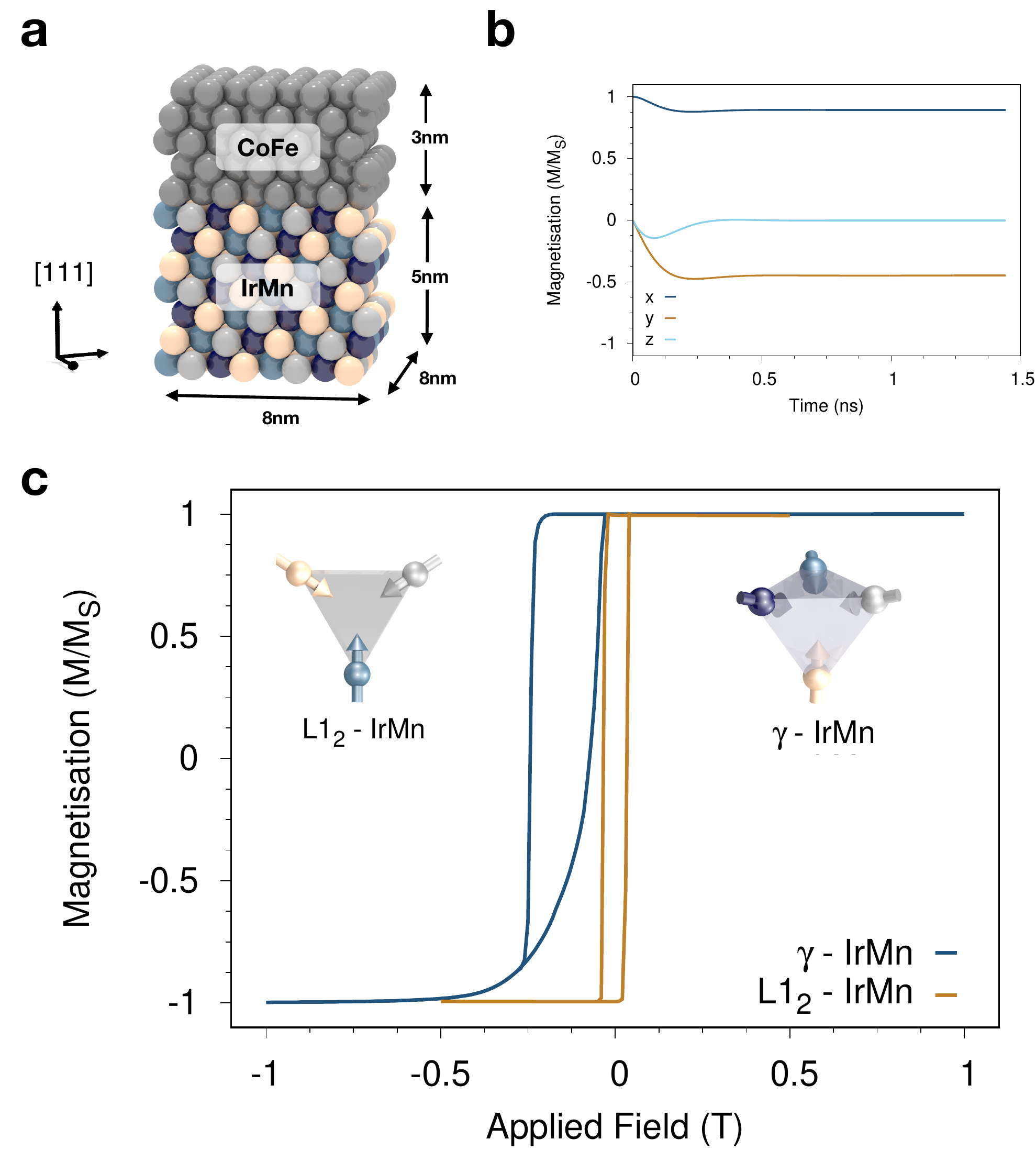}
\caption{\textbf{Schematic representation of the bilayer and simulated hysteresis loops.} (a) Schematic representation of the IrMn/CoFe bilayers. (b) The equilibration of the FM after  (c) \new{0K} hysteresis loops for the ordered and disordered phases of IrMn. \new{The magnetic structures of both $\gamma$ - \IrMn and $L1_2$ - \IrMn are shown within (c).}}
\label{fig:hy}
\end{figure}

\textit{Results} - To study the exchange bias effect, we couple the 5 nm thick \IrMn layer to a 3 nm thick ferromagnetic layer of CoFe to form a bilayer with a [111] out of the plane orientation of the \IrMn to reproduce the structure used in typical devices, as shown in Fig.~\ref{fig:hy}(a). \new{The system is comprised of 30,000 atoms in total.}

The direction of the exchange bias is set by a simulated field cooling process in a 0.1 T field applied along the $x$-direction. After cooling, the field is removed and the system relaxes to an equilibrium state. For the disordered $\gamma$-\IrMn, the CoFe magnetisation tilts around 19 degrees out of the plane due to imprinting from the underlying antiferromagnetic spin structure as shown in Fig.~\ref{fig:hy}(b).

We then simulate a hysteresis loop \new{at 0K} to calculate the exchange bias field, \new{the hysteresis loop is run for 40 ns between $\pm$ 1T in 0.01 T steps}. Fig.~\ref{fig:hy}(c) shows the simulated hysteresis loops comparing the ordered and disordered \IrMn phases. The magnetic field is applied along an axis parallel to the ferromagnet after equilibration to avoid spurious rotational effects. The loop for the disordered $\gamma$ -\IrMn /CoFe system shows an exchange bias field of $B_{\mathrm{EB}} = 0.14$ T.  Assuming a reduction in the exchange bias due to temperature effects, this value is close to typical experimental measurements~\cite{EBreview,HOhldag}. In stark contrast, the perfectly ordered \Lonetwo-\IrMn/CoFe system shows no exchange bias and very low coercivity with a completely symmetric loop. This system has no defects or lattice imperfections and therefore the exchange bias must be attributed to the intrinsic ordering, raising the question: how does the intrinsic ordering in the antiferromagnet determine the exchange bias? 

\begin{figure}[ht!]
\centering
\includegraphics[width= 0.45\textwidth]{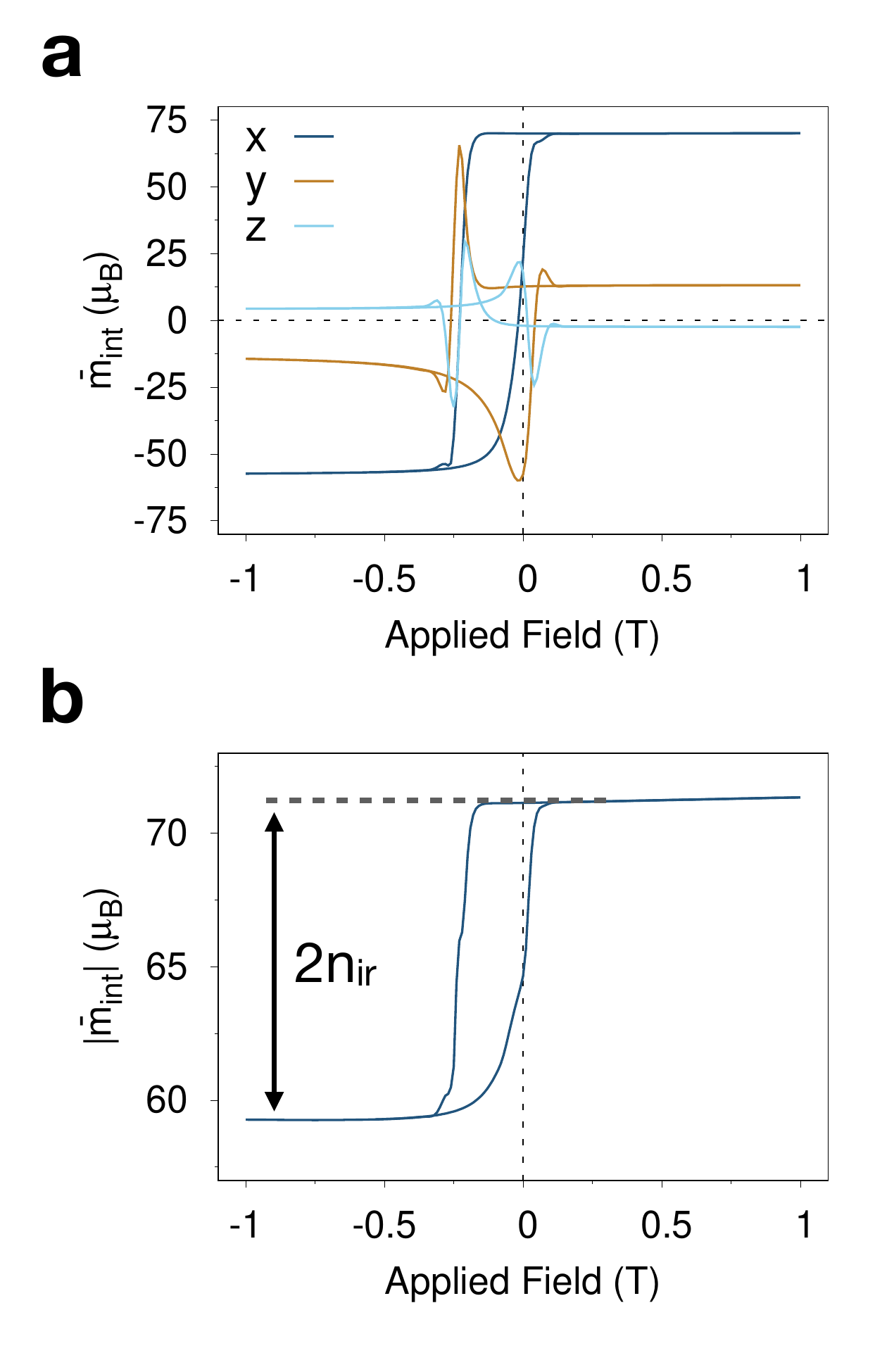}
\caption{\textbf{The interfacial origin of exchange bias through reversible and irreversible spins}. (a) Behaviour of the components of the net magnetic moment in the interfacial layer of $\gamma$ - \IrMn showing hysteretic behaviour of the interfacial moment which follows the CoFe magnetization, indicating a large reversible component of the magnetization. The loop is vertically shifted showing a change in the magnitude of the interfacial \IrMn moment during the hysteresis cycle (b), indicating the irreversible spins $n_{\mathrm{ir}}$ contributing to the exchange bias field.}
\label{fig:interface}
\end{figure} 

To determine the cause of exchange bias in the disordered $\gamma$-\IrMn CoFe bilayer, the direction and magnitude of the magnetisation of the interfacial AFM moments was analysed throughout the hysteresis loop, shown in Fig.~\ref{fig:interface}. The AFM interface moment exhibits a hysteresis loop with the same coercivity and exchange bias as the hysteresis loop of the FM, characteristic of a large reversible component of the interfacial magnetisation and in agreement with previous XMCD measurements~\cite{HOhldag}. The magnitude of the interface magnetisation exhibits a small vertical shift which is not present in the hysteresis loop of the FM. The change in magnitude arises due to the interface exchange field, where irreversible spins do not rotate during the hysteresis cycle. From this, we can conclude that our interface moment is comprised of a reversible moment  ($n_c$) and an irreversible moment ($n_{\mathrm{ir}}$). At positive saturation, the interface field is equal to $n_{+} = n_{\mathrm{r}} + n_{\mathrm{ir}}$, whereas at negative saturation the interface field is equal to $n_{-} = n_{\mathrm{r}} - n_{\mathrm{ir}}$. The vertical shift in the hysteresis loop is therefore equal to twice the number of uncompensated spins at the interface ($n_+ - n_-$). These uncompensated and irreversible spins are the spins which contribute to the exchange bias effect. In the example in Fig. \ref{fig:interface} the vertical shift is 13.92 atomic moments which corresponds to 6.96 uncompensated spins. \new{This is a very small percentage of the 16,000 total Mn atoms in the simulation}

The exchange bias is determined from the number of uncompensated irreversible spins as \cite{MBean}:
\begin{equation}
    |B_{\mathrm{EB}}| = \frac{n_{\mathrm{ir}}J_{int}}{\mu_{\mathrm{FM}}n_{\mathrm{FM}}}
    \label{eq:EB}
\end{equation}
where $n_{FM}$ is the number of ferromagnetic atoms and $\mu_{FM}$ is the magnetic moment of the FM atoms. Using $n_{\mathrm{ir}} = 6.96$ the exchange bias is calculated to be 0.13 T, in close agreement with the simulated value. 
The existence of irreversible spins is a direct output of our simulation and quantitatively correlates with the computed exchange bias field with a physically realistic magnitude. Importantly this is observed without the need for AFM grains or interface mixing. But what is the nature of these uncompensated spins? 

To this end we consider the nature of the atomic structure of disordered $\gamma$-\IrMn where 25\% of the atoms in each sublattice are Ir. The random distribution of the Ir atoms means that although on average 25\% are removed from each sublattice in reality a slightly different number will be removed when considering a finite number of interface spins. 
This statistical imbalance in the number of spins in each sublattice leaves a small net magnetic moment along the direction of the sublattice with the largest number of Mn atoms remaining. The number of atoms in each sublattice is shown in Tab.~\ref{table:EB} for the specific simulation in Fig.~\ref{fig:hy}.

\begin{table}[!tb]
\centering % used for centering table
\begin{ruledtabular}
\begin{tabular}{c c c c c } % centered columns (4 columns)
Sublattice & 1 & 2 & 3 & 4 \\
\hline % inserts single horizontal line
$N_S$      & 192  & 193  & 200  & 195  \\
\end{tabular}
\end{ruledtabular}
\caption[How the exchange bias is predicted from the crystallography.]{The number of atoms $n_l$ in each magnetic sublattice $l$. This gives $n_{\mathrm{ir}}$ as 6.67 uncompensated spins, as calculated from Eq.~\ref{eq:un}. This imbalance is caused by there being an average of 6.67 atoms more in sublattice 3 than in the other 3 sublattices, while the magnitude is reduced due to sublattice disorder arising from local spin frustration.
} % title of Table
\label{table:EB} % is used to refer this table in the text
\end{table}

From the number of atoms in each sublattice the number of irreversible interface spins can be calculated by 
\begin{equation}
    n_{\mathrm{ir}} = n_{\mathrm{max}} - n_{\mathrm{av}},
    \label{eq:un}
\end{equation}
where $n_{\mathrm{max}}$ is the number of Mn atoms in the sublattice with the largest number of atoms and $n_{\mathrm{av}}$ is the average number of Mn atoms over the other three sublattices. The calculation gives $n_{\mathrm{ir}} = 200 - 193\tfrac{1}{3} = 6.67$ for our interface, an almost exact match to the result calculated in Fig.~\ref{fig:interface}. The discrepancy between the values is because our calculation is simplified and assumes that the bias field lies exactly along the direction of one of the AFM sublattices. In reality this will only occur when $n_{\mathrm{max}} >> n_{\mathrm{av}}$ and the direction of the bias field will be a vector combination of all four sublattice magnetisation directions which depends on the positions of the removed atoms. The fraction of irreversible spins is only 0.9\% of the total interface atoms. This small imbalance combined with a large exchange interaction predicts an exchange bias field of 0.15 T using Eq.~\ref{eq:EB} which is very close to the numerical simulation of 0.14 T.

In ordered \IrMn, the Ir atoms are not randomly located and are instead all removed from the same sublattice leaving a perfectly compensated spin structure as found experimentally~\cite{kohn}. In this structure there are no irreversible spins, which means the AFM interface has no net interface magnetisation, explaining the absence of exchange bias in the simulation. \new{Here the grain size has been kept constant to 8nm $\times$ 8nm, but we expect that as the grain size is reduced the exchange bias should increase due to the increased statistical imbalance but decreasing at finite temperatures due to reduced thermal stability. Grain size effects in exchange biased systems will be discussed in a future paper.}

%The simulated system is a very simplified case due to the atomically flat interface. In reality, the interface will not be atomically flat and this disorder may cause the sample to exhibit some exchange bias. 

\begin{figure}[tb!]
\centering
\includegraphics[width= 0.45\textwidth]{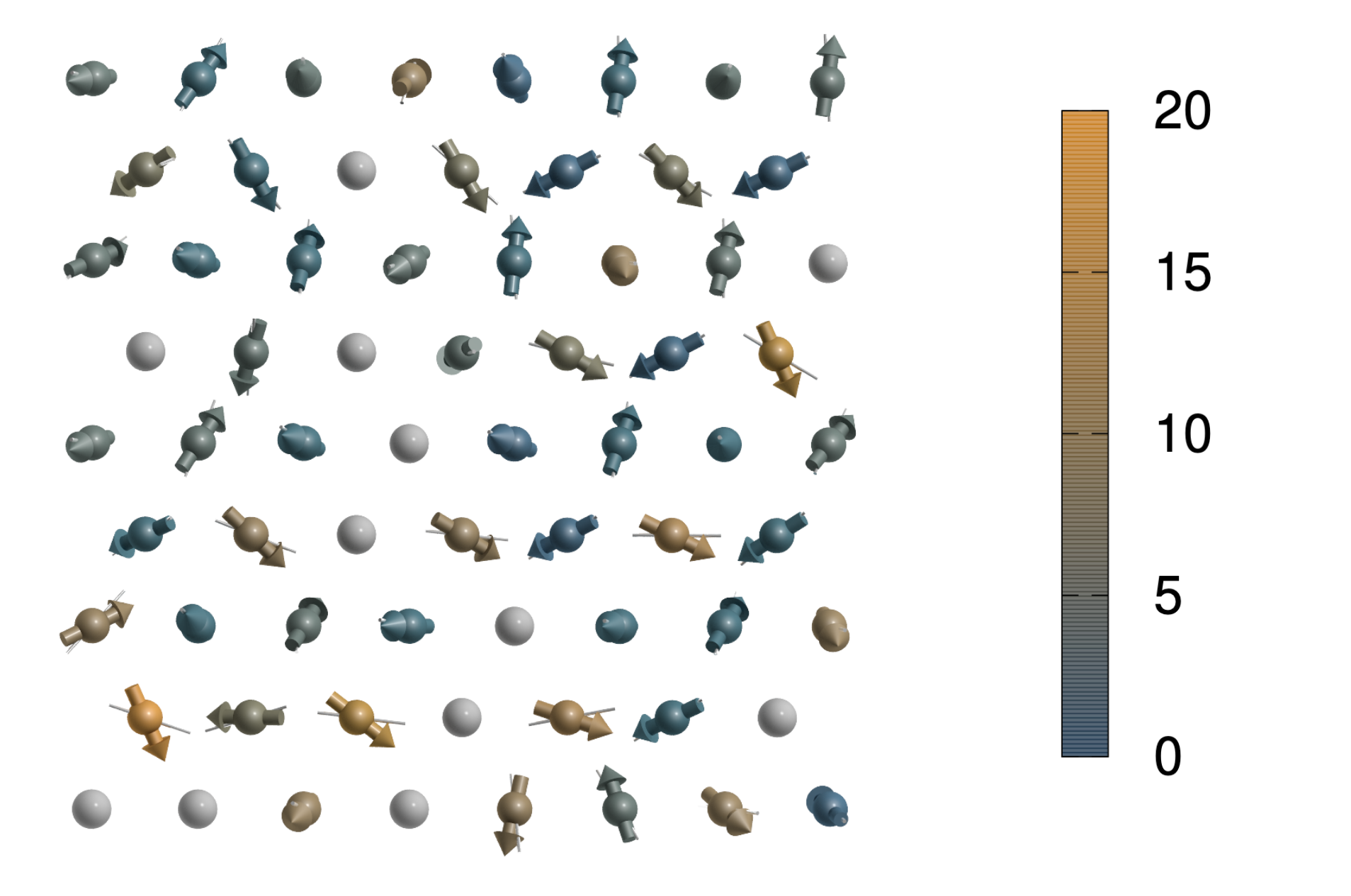}
\caption[A visualisation of the net magnetic moment of the interface layer throughout the hysteresis loop]{\textbf{A visualisation of the net magnetic moment of the interface layer throughout the hysteresis loop.} Arrows indicate spin positions at negative saturation, thin bars indicate the starting positions at positive saturation and the colour scale shows the angular change. The small angular deviation of individual spins demonstrates the delocalized nature of the reversible and irreversible spins.}
\label{fig:interface_motion}
\end{figure}

Now that we have identified the cause of the irreversible spins in the disordered $\gamma$ - \IrMn, we want to know where they are located across the interface. There are two options for the location of these spins: (a) 6-7 specific spins are pinned and the rest reverse as normal or (b) a small proportion of every spin is pinned (each spin is approximately ~9\% pinned). For the rest of the paper option (a) is referred to as localised pinning and option and option (b) as delocalized pinning. To work out whether the pinned spins are localised or delocalized, the interface spin structure was visualised throughout the hysteresis loop. A visualisation of a small section of the interface is shown in Fig.~ \ref{fig:interface_motion}. The direction of magnetisation of each atomic spin was compared between positive saturation and negative saturation points in the hysteresis loop. Our simulation reveals that each of the interfacial spins moves only slightly between positive and negative saturation, amounting to a small distortion of the interfacial spin structure. The irreversible spins come from a net change in the total interfacial moment delocalized across the interface rather than the  reversal of individual spins, suggesting the pinned spins are delocalized. The strong exchange coupling between the spins stabilises the overall spin structure preventing a large angular change for individual spins.

\begin{figure}[tb!]
\centering
\includegraphics[width= 0.45\textwidth]{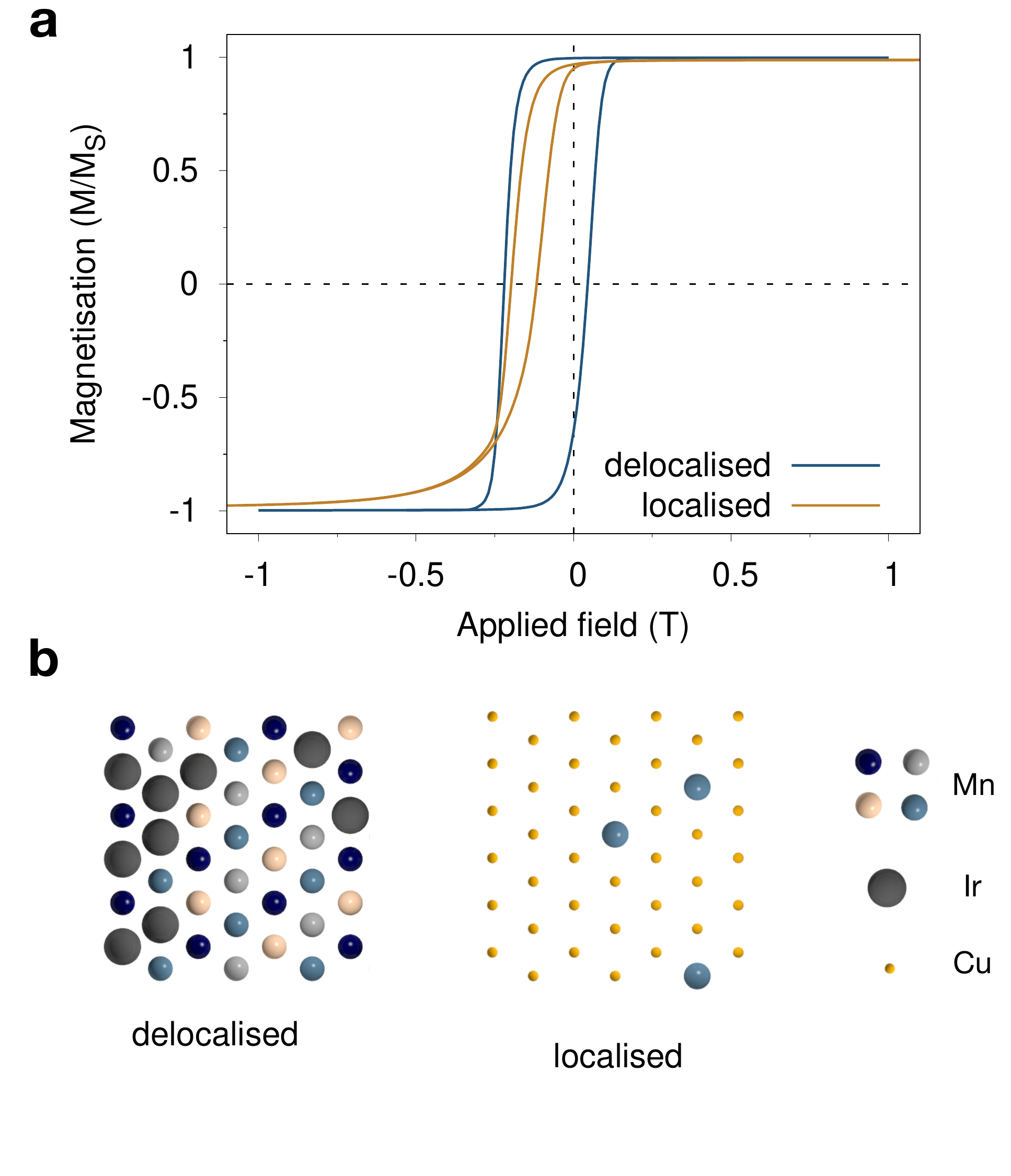}
\caption{\textbf{The difference between localised and delocalized spins} (a) Exchange bias simulations for localised and delocalized uncompensated spins. (b) The difference in atomic structure in the interface AFM layer between delocalized distributed and localised spins at the interface.}
\label{fig:delocal}
\end{figure}

Previous models of exchange bias have focused on the irreversible interface spins being due to localised spins~\cite{domainstate}. Our model shows that in fact the exchange bias effect can be caused by delocalized spins. We suggest that although delocalized spins do cause exchange bias, localised spins can also cause exchange bias. In real devices, localised spins would occur due to defects such as point defects, non magnetic impurities or grain boundaries. To confirm this we have simulated a  disordered - $\gamma$ -\IrMn\ bilayer but without the delocalized interface spins by removing all of the AFM interface spins apart from 6 taken from the same sublattice, which are still coupled to the bulk AFM and to the FM. These are localised point spins and are the only ones which contribute to the exchange bias field. The rest of the interface spins are replaced with non-magnetic Cu atoms. The hysteresis loop produced is shown in Fig. \ref{fig:delocal} and is compared to our previous hysteresis loop which occurred due to the delocalized interface spins. Both of these simulations exhibit similar levels of exchange bias which is stable with temperature. For localised spins all the spins in the interface layer contribute to the exchange bias field. 
%\cmr{What do the interface spins do here? presumably they have no or a small reversible component? Could this be a way to massively increase the exchange bias? - ie by fudging the statistics?} 
One intriguing possibility is using statistics to engineer the magnitude of the exchange bias field by Cu dusting. In reality the Mn atoms will fall randomly on the different magnetic sublattices. However, the statistical imbalance of spins is larger for smaller interfaces, thereby increasing the exchange bias field, as seen experimentally~\cite{VinaiAPL2014}.

\textit{Conclusion} - We have determined the microscopic physical origin of exchange bias in the complex IrMn/CoFe system, a long standing problem in the fields of magnetism and spintronics. Exchange bias was found to exist in our perfect FM/AFM bilayer without the need for domain walls or impurities. It was found that exchange bias is caused by a small statistical imbalance in the number of Mn atoms in each AFM sublattice. The imbalance causes a net field at the interface which pins the FM, causing exchange bias and explaining the observation of exchange bias in disordered IrMn. There is no observable exchange bias in ordered IrMn as there are no uncompensated spins \new{for the (111) orientation, while some bias can occur for the (001) interface} \cite{kohn}. In our simulations it was found that the uncompensated spins were delocalized across the interface, however, exchange bias can still occur in systems with localised uncompensated spins. Our findings provide new insight into the physical origin of exchange anisotropy by accounting for the correct nature of the antiferromagnetic spin structure and crystallography, finally resolving one of the most complex and outstanding challenges in the field. The enhanced understanding will provide new routes for optimisation of nanoscale exchange biased systems with relevance to upcoming neuromorphic~\cite{GrollierNelec2020} and antiferromagnetic spintronic \cite{BaltzRMP2018,JungwirthNnano2016} devices.

\bibliography{EB.bib}

\end{document}